\documentclass[aps,pre,twocolumn,superscriptaddress,floatfix,amsmath,amssymb,showpacs,preprintnumbers, 10pt]{revtex4-2}
\usepackage{amsmath,amsfonts,amssymb,amsthm,graphics,graphicx,epsfig,bbm}
\usepackage[colorlinks=true,citecolor=blue,linkcolor=blue,urlcolor=blue]{hyperref}
\usepackage[usenames]{color}
\usepackage[T1]{fontenc}
\usepackage{lmodern}
\usepackage{graphicx}
\usepackage{subfigure}
\usepackage{amsmath}
\usepackage{epsfig}
\usepackage{dcolumn}
\usepackage{bm}
\usepackage{color}
\usepackage{times}
\usepackage{epstopdf}
\usepackage{amssymb}
\usepackage{lipsum}
\usepackage{amstext}
\usepackage{latexsym}
\usepackage{amsfonts}
\usepackage{psfrag}
\usepackage{dsfont}
\usepackage{braket}
\usepackage{comment}
\usepackage{booktabs}
\usepackage{float}

\usepackage[nameinlink]{cleveref}
\crefname{equation}{Eq.}{}
\Crefname{equation}{Eqs.}{}
\crefname{figure}{Figs.}{Figs.}
\Crefname{figure}{Fig.}{Fig.}

\crefname{table}{Table}{Table}
\crefname{Appendix}{Appendix}{Appendices}

\usepackage{lineno}

\begin{document}


\title{Emergent collective dynamics from motile photokinetic organisms}

\date{\today}

\author{J. Morales}
\affiliation{Departamento de Matem\'atica, Universidad Tecnol\'ogica Metropolitana, Santiago, Chile}

\author{P. Mu\~noz}
\affiliation{Departamento de Matem\'atica, Universidad T\'ecnica Federico Santa Mar\'ia, Santiago, Chile}

\author{D. Noto}
\affiliation{Department of Earth and Environmental Science, University of Pennsylvania, Philadelphia, USA}

\author{H. N. Ulloa}
\email{ulloa@sas.upenn.edu}
\affiliation{Department of Earth and Environmental Science, University of Pennsylvania, Philadelphia, USA}

\author{F. Guzm\'an-Lastra}
\email{fguzman@uchile.cl}
\affiliation{Departamento de F\'isica, Facultad de Ciencias, Universidad de Chile, Santiago ,Chile}


\begin{abstract}
The day–night cycle drives the largest biomass migration on Earth: the diel vertical migration (DVM) of aquatic organisms. Here, we present a three-dimensional agent-based model that incorporates photokinesis, gyrotaxis, and stochastic reorientation to explore how individual-level swimming behaviors give rise to population-scale DVM patterns. By solving Langevin equations for swarms of swimmers, we identify four distinct regimes—{\it Surface Accumulation}, {\it Shallow DVM}, {\it Deep DVM}, and {\it Sinking}—governed by two key dimensionless parameters: the Péclet number ($Pe$), representing motility persistence, and the vertical swimming asymmetry ratio ($W = w_{\text{down}} / w_{\text{up}}$), encoding photokinetic bias. These regimes emerge from nonlinear interactions between light-driven navigation and active noise, diagnosed through topological and statistical features of vertical distributions. A critical feedback is uncovered: upward-biased swimming ($W < 1$) promotes surface aggregation, while excessive downward bias ($W > 1$) leads to irreversible sinking. Analytical estimates link regime boundaries to gyrotactic alignment and velocity reversals. Together, our results provide a mechanistic framework to interpret DVM diversity and emphasize the central role of light gradients—beyond absolute intensity—in shaping ecological self-organization.
\end{abstract}

\flushbottom
\maketitle
%
%
\thispagestyle{empty}


The diurnal cycle of sunlight drives the largest coordinated swimming of organisms in Earth's aquatic ecosystems \cite{hays2003review,bianchi2013diel}. Each day, swarms of motile plankton ascend at dusk and descend at dawn, forming a global-scale biological wave that shapes biogeochemical processes, trophic dynamics and climate \cite{bianchi2013diel,steinberg2017zooplankton,archibald2019modeling}. This phenomenon—known as diel vertical migration (DVM) \cite{ringelberg2010diel,bandara2021two}—is primarily governed by light cues and reflects an evolutionary trade-off between predator avoidance, nutrient acquisition, and photosynthetic efficiency \cite{gliwicz1986predation,bollens1989predator,lampert1989adaptive,neill1990induced,kelly2019importance,ludvigsen2018use}. At the organismal level, DVM arises from the interplay of phototaxis, photokinesis, buoyancy regulation, and the power of plankton to swim at speeds up to ten body lengths per second \cite{fenchel2002microbial,jekely2009evolution,noss2014direct,dean2016biomixing,ouillon2020active,hafker2022animal,hafker2023rhythms,zeng2022sharp,noto2023simple}. These microscale behaviors self-organize into large-scale patterns, further modulated by stratification, shear, and convection \cite{pedley1992hydrodynamic,durham2012thin,chen2021diel,sengupta2017phytoplankton,michalec2017zooplankton,houghton2018vertically,bees2020advances,wang2021vertical,perga2020rotiferan}. Despite over two centuries of research, the mechanisms driving the diversity of collective DVM behaviors—ranging shallow to deep excursions and tightly clustered to dispersed distributions—remain poorly understood, hindering our ability to predict how organisms respond to changing environmental conditions \cite{bandara2021two}.

To capture the multiscale dynamics of collective swimming, two principal modeling frameworks have been developed: continuum models \cite{richards1996diel,han1998modeling,ji2007vertical,bianchi2013diel} and agent-based models \cite{ouellette2022physics,countryman2022modelling,huang2024collective,hang2025self}. Continuum models treat organism density as a continuous field governed by advection–diffusion equations \cite{huang2023complex,potts2021modelling,bees1998linear,richards1996diel}, where phototactic behavior is represented by drift terms and dispersion by turbulence-dependent diffusivity \cite{arboleda2023self,kessler1986individual,arrieta2019light,javadi2020photo,vincent1996bioconvection,pedley1990new}. While these models offer analytical tractability and capture large-scale DVM patterns, they often neglect key microscale mechanisms such as adaptive motility and signal integration, limiting their capacity to link behavior to emergent structure. Agent-based models take a ``bottom-up'' approach \cite{ouellette2022physics}, explicitly simulating the Lagrangian trajectories of self-propelled agents that respond to local light cues via phototaxis, photokinetic speed modulation, stochastic fluctuations, and circadian switching \cite{bearon2006relating,son2013bacteria,bianchi2013diel,chen2021diel,leptos2023phototaxis,dervaux2017light,houghton2018vertically}. These models reveal how sensory-motor coupling and environmental feedback drive complex, emergent dynamics. Yet, despite their mechanistic richness, existing agent-based models have not quantitatively bridged organism-scale behavior with population-scale migration patterns \cite{ouellette2022physics,countryman2022modelling,monthiller2022surfing}. Key questions remain: What determines the vertical extent and the spatial distribution of migratory swarms? \cite{bees1998linear,kessler1986individual,krishnamurthy2020scale,hafker2023rhythms,son2013bacteria} And crucially, can physics-based, individual-scale models predict the spatiotemporal organization observed in natural systems?

Here, we introduce an agent-based model (ABM) framework and present numerical simulations of motile organisms encoded with photokinesis, gyrotaxis, and Brownian motion. These individual-scale behaviors give rise to a constellation of self-organized, population-level patterns. The emergence of these patterns is regulated by spatiotemporal variations in light and captured through two key dimensionless parameters: the Péclet number ($Pe$), quantifying the balance between directed motion and rotational diffusion, and the vertical swimming asymmetry ratio ($W$), which measures vertical motility bias \cite{schuech2014going}. Our results uncover a previously unrecognized role for the interplay between photokinetic speed modulation and gyrotactic alignment in shaping both individual trajectories and emergent DVM structures. Our findings not only advance the mechanistic understanding of ecological self-organization in light-sensitive systems but also provide a predictive framework for designing and controlling phototactic active matter.\\

\begin{figure*}[ht!]
    \centering 
\includegraphics[width=\textwidth]{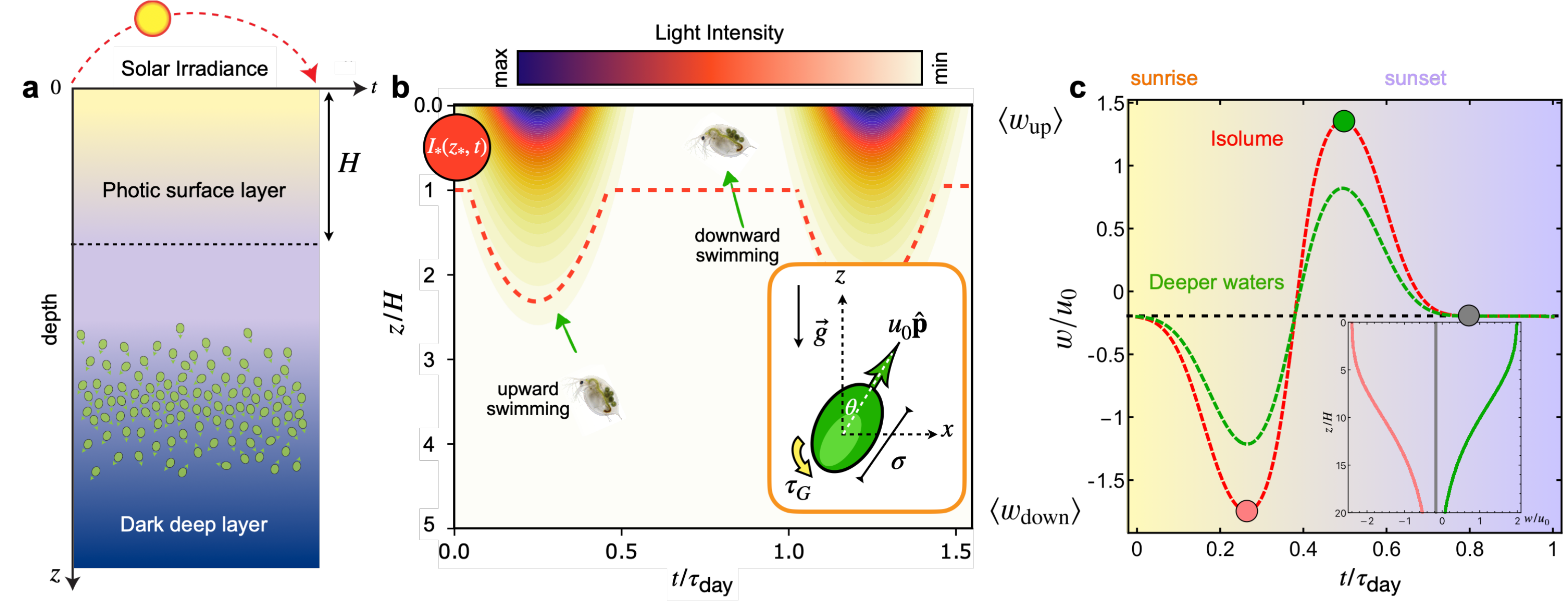}	
    \caption{\textbf{Diel vertical migration of motile photokinetic organisms}. (a) Scheme of a colony of zooplankton homogeneously distributed along the water column $L_z$, with light intensity decreasing exponentially with depth $z$. The depth $H$ denotes the niche region where organisms tend to cluster during nighttime. (b) Light intensity model $I(z,t)$. Light intensity is maximum at the surface of the water column and decreases with depth, it also varies with day cycle, reaching a minimum at night. Zooplankton communities such as {\it Daphnia magna} follows a light intensity niche called the isolume $z_*$ represented in a red dashed line. Inset: Zooplankton minimum model, a spherical particle of diameter $\sigma$ moves with velocity $u_0$ and director vector $\hat{\mathbf{p}}(t)$, the swimmer is subject to gyrotactic response $\tau_G$ and rotational Brownian motion $D_R$. (c) Swimming strategy vertical velocity. In red dashed line is shown the velocity changes during a day cycle for a swimmer at the isolume $z=z_*$, in green dashed line the velocity changes for a swimmer at deeper waters $z>z_*$. Inset: Velocity changes with depth at three specific day times, in pink close to midday, in green during sunset, and in gray at night.} 
\label{fig1}
\end{figure*}

\noindent{\fontsize{12}{14}{\textbf{Results}}}

\noindent\textbf{ABM Modeling framework}

\noindent Light is widely recognized as the primary environmental cue regulating the circadian vertical migration of motile organisms in both marine and lacustrine systems \cite{ringelberg2010diel}. During daylight hours, many species position themselves at depths corresponding to a ``dusky-light niche'', defined by a critical isolume intensity $I_*$ located at a time-dependent depth $z_{*}(t)$ \cite{javadi2020photo,richards1996diel,hafker2023rhythms}. At night, these organisms ascend toward the upper water column or aggregate near the thermocline, where larger zooplankton typically graze on smaller phytoplankton. Vertical migration is thus tightly coupled to light availability, and organisms respond rapidly to changes in solar and lunar illumination \cite{sherman1970vertical,last2016moonlight}, particularly when other light signals are weak or absent. In our model, a swarm of motile organisms follows a swimming strategy modulated by a circadian spatiotemporal light field, in which intensity decreases with depth, and varies between day and night (Fig.~\ref{fig1}). Each organism is treated as an active photokinetic agent of size $\sigma$ that self-propels at a characteristic speed $u_{0}$ and adjusts its vertical velocity in to maintain alignment with the isolume $I_*(t)$~\cite{desai2017modeling,leptos2023phototaxis}, as illustrated in Fig.~\ref{fig1}b. Their swimming direction is influenced by rotational diffusion and a gyrotactic torque, which reorients the body against gravity over a characteristic timescale $\tau_G$ (inset Fig.~\ref{fig1}b). Our modeling framework does not allow spatial overlap of organisms, meaning they need to reorient and move to avoid collisions while swimming (\textcolor{blue}{Methods}).

Light intensity in the water column varies with both depth and time and is modeled as
\begin{equation}\label{eq:light_model}
I(z,t)=I_{0}\zeta(z)\,\tau(t),
\end{equation}
where $I_{0}$ is the light intensity at the surface ($z=0$), $\zeta(z)=\operatorname*{exp}(-k_d z)$ describes the vertical attenuation following the Beer--Lambert law, with $k_d$ denoting the bulk light attenuation coefficient \cite{richards1996diel,ringelberg2010diel}. Temporal variations are captured by a smooth function representing a canonical day-night cycle: $\tau(t)=\tau_0\left[1-\sin \left(2 \pi t/\tau_{\rm day}-\phi\right)\right]^b$, with model parameters $\tau_0=1/16$, $b=4$ and $\phi = \tau_{\rm day}/2$. To stay within their preferred light niche, swimming organisms continuously sense the local light intensity $I(z,t)$. By comparing it to their preferred light level, or isolume $I_*(t)$, they determine whether they need to move upward or downward in the water column. This comparison sets a target depth $z_*(t)$, which guides their vertical movement: 
\begin{equation}
 z_{*}(t)=
 \left\{\begin{array}{ll}
        \dfrac{1}{k_{d}}\log\left[\dfrac{I_{0}}{I_{*}}\tau(t)\right], & \text{for daytime: } I>0\\
        H, & \text{for nighttime: } I=0.
        \end{array}\right.
\end{equation}
Thus, during daylight, organisms pursue a time-varying depth where light matches their preferred range. At night, they swim toward a constant depth near the surface, representing their typical nocturnal niche \cite{ringelberg2010diel}.
If an organism is above this depth, it dives at a rate $w_{\rm down}$ until it reaches or falls below the isolume $I_*$. A photokinetic response then follows. From sunrise to midday—when the isolume $z_*(t)$ deepens—organism swims downward at a nondimensional speed $w_{\rm down}/u_0$. After midday, as the isolume ascends, the swimmer reverses direction and ascends at $w_{\rm up}/u_0$ until sunset. During nighttime, light-guided vertical migration ceases $\langle w\rangle /u_{0} \approx 0$, (with $\langle\cdot\rangle$ a time- and horizontally-averaged operator, \textcolor{blue}{Methods}) and organisms remain statistically at fixed depth $H$ until the next dawn begins. As an example, Fig.~\ref{fig1}c shows the swimming strategy of organisms that swim faster downward than upward \cite{bianchi2016global}. Green trajectories represent motion when an organism depth $z_{\rm org}$ is deeper than the target depth $z_{*}$, whereas red trajectories correspond to the case $z_{\rm org} = z_{*}$ (organisms are about the niche isolume). The inset in Fig.~\ref{fig1}c highlights depth-dependent velocity changes with pink, green, and gray curves, corresponding to specific times marked by colored circles in the main panel. These trajectories illustrate how the ability of organisms to track their preferred isolume $I_{*}$ depends on both their propulsion speed and position $z_{\rm org}$ relative to $z_*(t)$.

The swimming strategy is described by the following mathematical model:
\begin{equation}
w(z, t) = 
\begin{cases} 
-w_{\rm down} & \text{if } I \geq I_* \\ 
\\
\frac{1}{2}w_{\rm up}\left[1 + \operatorname{tanh}(A_1) \right] + \\
\frac{1}{2}w_{\rm down}\left[2\operatorname{tanh}(A_2) -1\right]  & \text{if } I < I_*
\end{cases}
\label{swimst}
\end{equation}
where $A_1=-[S(z,t)+R] /L$ and $ A_2=-(S(z,t)-R) /L$. The parameter $L$ quantifies the sensitivity of zooplankton to light stimuli: small values ($L\ll 1$) produce sharp responses, while larger values ($L\gg 1$) yield smoother transitions (\textcolor{blue}{Supplementary Fig.~S1}). The threshold parameter $R$ represents the rheobase \cite{richards1996diel}, with typical values $R<20$; for $R\geq 20$, the swimming response is strongly attenuated, resulting in near-zero, $w(z,t)=0$. The functional $S(z,t)$ captures the local rate of change in light intensity and is defined as $S(z,t)=\partial I(z,t)/\partial t+k_d$, where $I_0$ is the maximum light intensity at the water surface ($z=0$).\\

\noindent\textbf{Brownian versus ABM excursions}

\begin{figure}[h!]
    \centering 
\includegraphics[width=0.9\columnwidth]{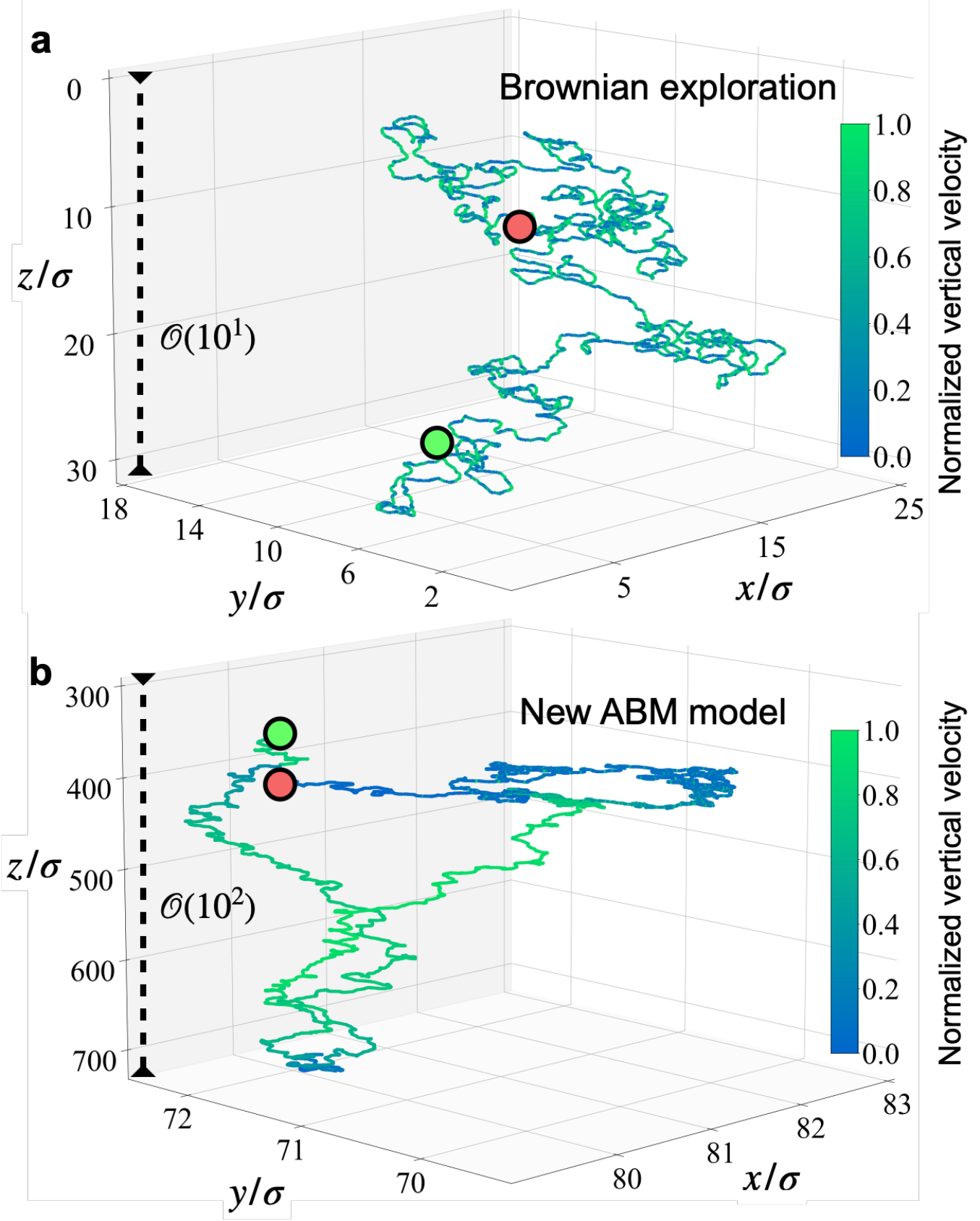}	
    \caption{\textbf{Trajectory and vertical swimming speed of a single organism}. Simulated trajectories and normalized vertical-component swimming speeds ($|w/u_{0}|$) of an organism encoded with photokinesis, gyrotaxis, and Brownian behaviors under two light regimes: (a), constant darkness throughout a diel cycle; (b), a day–night light cycle in which photokinesis is activated during the illuminated phase. The green dot represents their initial position and the pink dot their final position.} 
\label{fig2}
\end{figure}

\noindent Let's first examine the behavior of an individual organism. A first consequence of light-mediated swimming is a substantial increase in vertical excursion compared to purely Brownian exploration. To illustrate this, we compare the trajectories of two self-propelled spherical organisms of body length (diameter) $\sigma$: one exposed to constant darkness ($I = 0$) and the other to a day-night light cycle ($I > 0$). Their paths are shown in Fig.~\ref{fig2}, where green and red dots mark initial and final positions after one cycle, respectively. In darkness, the organism exhibits a noisy, spiral trajectory~\cite{thorn2010transport} driven by rotational diffusion ($D_R = 10\,\tau_G^{-1}$) and gyrotaxis, resulting in an exploration length scale $\ell_{\rm e} \propto \mathcal{O}(10^1\sigma)$ and a vertical speed uniformly distributed between 0 and 1 (Fig.~\ref{fig2}a). In contrast, under a day-night light cycle, the same organism undergoes a markedly different motion (Fig.~\ref{fig2}b), characterized by a long-range back-and-forth vertical excursion on the order of $\ell_{e} \propto \mathcal{O}(10^3\sigma)$ and a bi-modal swimming speed distribution. These contrasting trajectories highlight the powerful role of light and photokinesis in enhancing motility, enabling organisms to explore their environment across vastly larger spatial scales than possible through stochastic dynamics alone.\\

\noindent\textbf{Population-scale collective patterns}

\noindent 
Upscaling the analysis to a swarm of swimming organisms reveals the emergence of population-scale collective dynamics. Initially, organisms are uniformly distributed in the $x$-$y$ plane, and their vertical initial position is uniformly distributed around the nocturnal niche $z=H$ (\textcolor{blue}{Supplementary~Fig.~S2}). Then, their subsequent trajectories evolve according to the Langevin equations of motion (\textcolor{blue}{Methods}). From individual trajectory, we zoom-out and reconstruct the spatiotemporal concentration field of the swarm over four diel light cycles and investigate the resulting macroscopic behavior in terms of two key parameters. The first is the organism-scale Péclet number, $Pe = u_0 / (D_R \sigma)$, which quantifies the ratio between active self-propulsion and rotational diffusion. For $Pe < 1$, the swimmer's motion is predominantly diffusive, whereas for $Pe > 1$, directional persistence dominates. The second parameter is the vertical speed ratio during the diel vertical migration, defined as $W = w_{\rm down}/w_{\rm up}$. Without loss of generality, and to reduce computational cost, we define the niche isolume as the time-dependent depth at which light intensity reaches $85\%$ of the surface maximum, $I_{*}=0.85I_0$. Simulations are performed in a three-dimensional domain of depth $L_z=4\times 10^3\sigma$, where light intensity in the deepest region is approximately $5\%$ of the surface value (\textcolor{blue}{Supplementary~Fig.~S1}). All organisms and simulations have the same characteristic self-propulsion speed $u_0$, gyrotactic reorientation timescale $\tau_G=1$, and upward swimming speed $w_{\rm up}=2u_{0}$. To explore the population-scale response, we systematically vary $Pe$ and $W$ by tuning the rotational diffusion coefficient $D_R$ and downward swimming speed $w_{\rm down}$. These parameter ranges encompass behaviors observed across a wide variety of marine microorganisms~\cite{vladimirov2004measurement,krishnamurthy2020scale,sengupta2017phytoplankton}. A complete list of simulation parameters is provided in \textcolor{blue}{Supplementary~Table~S1}.

\begin{figure}[H]
    \centering 
\includegraphics[width=\columnwidth]{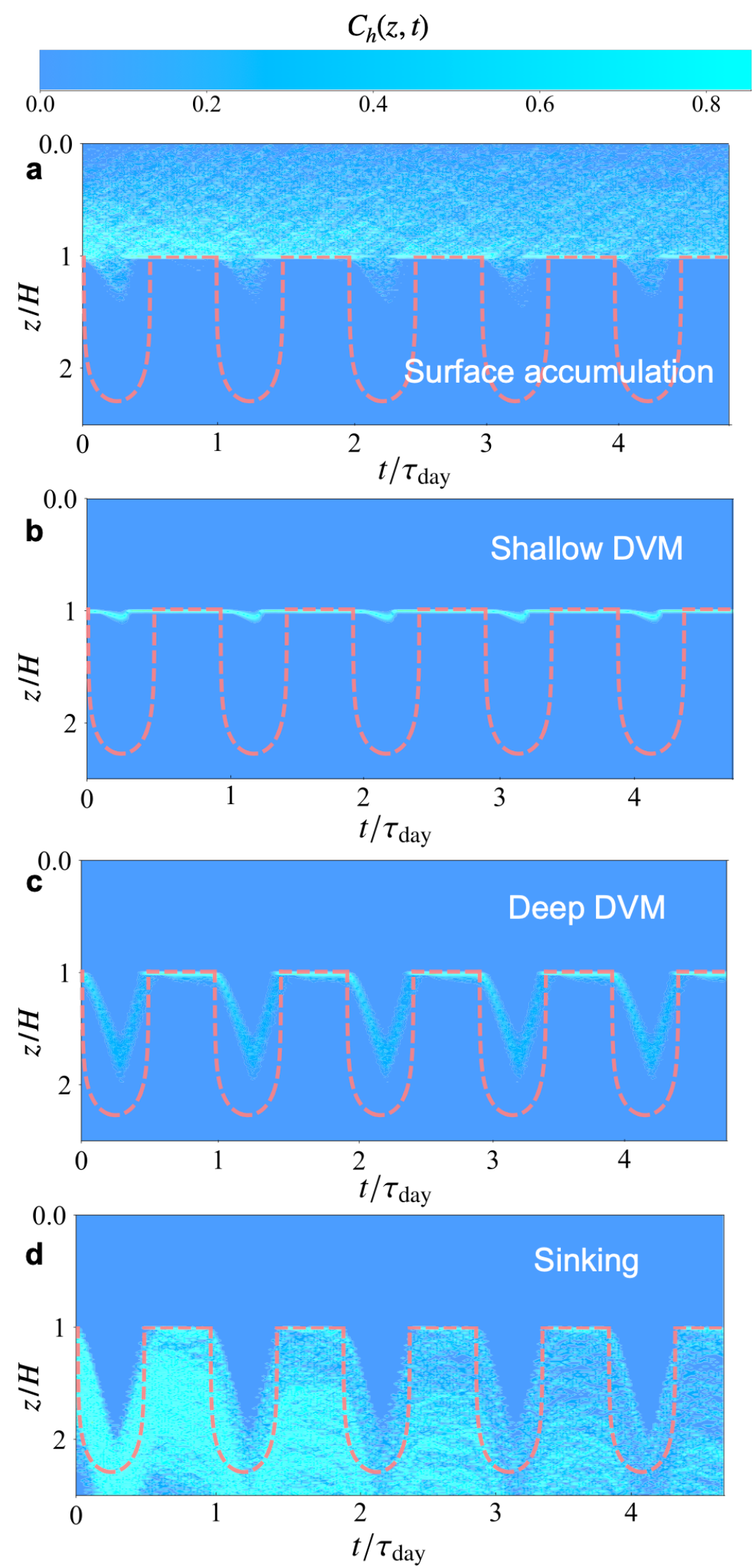}	
\caption{\textbf{Swarm collective patterns under diel light forcing.}
Horizontally averaged vertical concentration $C_{h}(z,t)$ over four diel cycles for key combinations of Péclet number and vertical speed ratio, highlighting distinct collective behaviors. Dashed curve highlights the time-dependent dimensionless depth $z_{*}(t)/H$ of the target isolume $I_{*}$ during daytime and the noctunal niche $z/H=1$ during nighttime.
(a) {\it Surface Accumulation} ($Pe = 100$, $W = 0.1$): swarm confined to the upper 10\% of the water column.
(b) {\it Shallow DVM} ($Pe = 0.1$, $W = 0.1$): thin layer forms at the nocturnal niche ($z/H = 1$).
(\textbf{c}) {\it Deep DVM} ($Pe = 10$, $W = 0.8$): synchronized tracking of the diel isolume.
(d) {\it Sinking} ($Pe = 10$, $W = 1.2$): gradual descent and dispersion below the niche isolume.} 
\label{fig3}
\end{figure}

To quantify how individual-scale swimming strategies shape population-level distributions, we computed the probability density function (PDF) $n(z)$ of the time- and horizontally-averaged concentration field $C({\bm x},t)$, which captures the statistics of the spatiotemporal distribution of swimmers. This metric parallels DVM patterns commonly inferred from backscatter signals in Acoustic Doppler Current Profiler (ADCP) measurements \cite{lorke2004acoustic,cheslack2023diel,piton2025identifying}. We examined four representative combinations of the Péclet number and vertical speed ratio, each corresponding to a distinct dynamical regime. These cases reveal how differences in self-propulsion and vertical asymmetry drive transitions from diffuse dispersal to sharp stratification, governed by persistent, light-sensitive swimming. The resulting vertical and temporal distribution of the organisms' concentration $C_{h}(z,t)$ (Fig.~\ref{fig3}, \textcolor{blue}{Methods}) highlights the diversity of collective behaviors emerging from simple local rules under diel light forcing. Although all the collective patterns demonstrate DVM, each of them presents specific features that we highlight and characterize next.

\begin{figure}[ht!]
    \centering 
\includegraphics[width=\columnwidth]{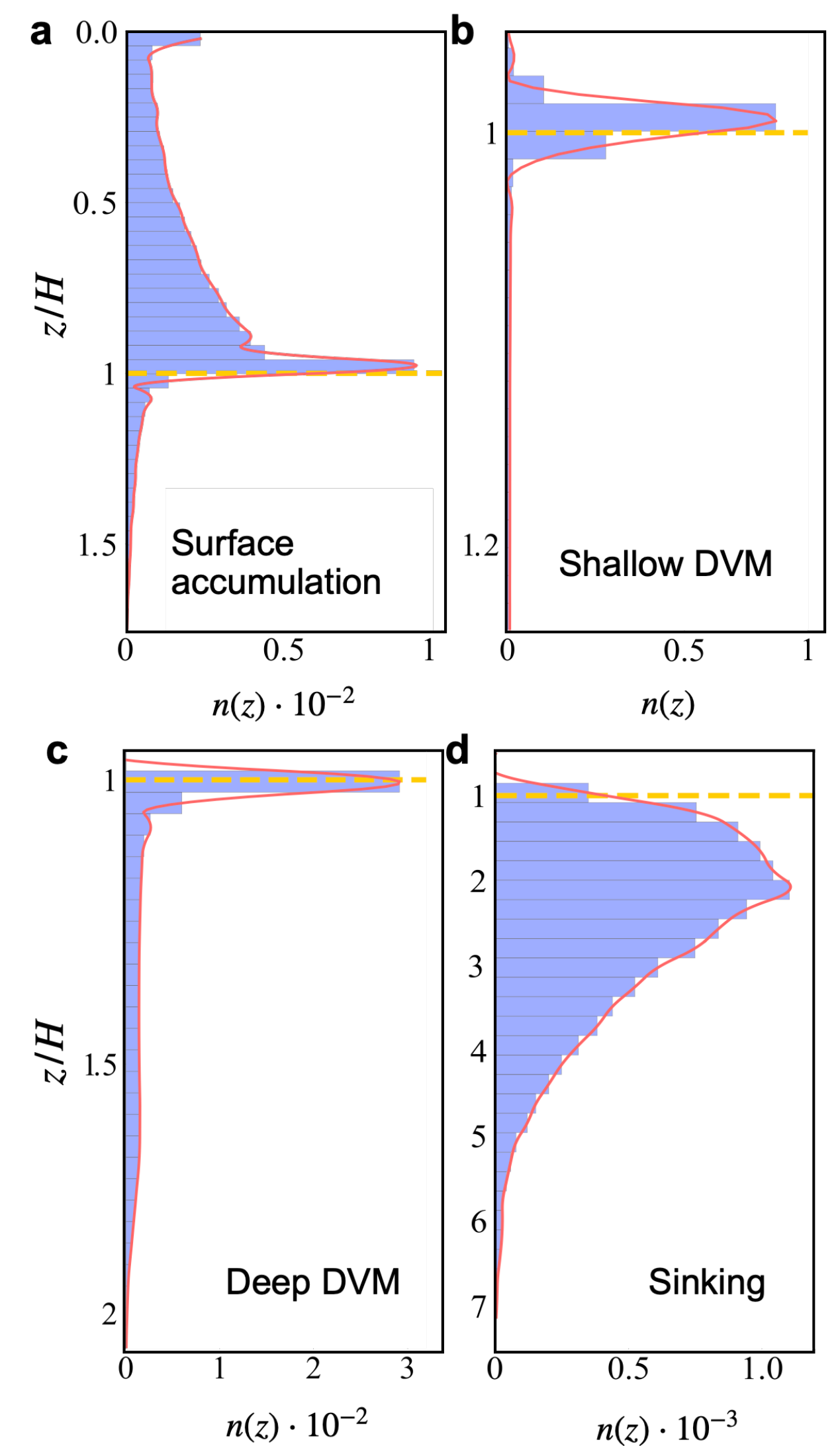}	
    \caption{\textbf{Vertical distribution of organisms across migration regimes}.
Probability density distribution $n(z)$ of the time- and horizontally-averaged concentration, $\langle C({\bm x},t) \rangle$, over a four-day cycle for representative combinations of simulation parameters, illustrating the four identified behavioral regimes. The yellow dashed line marks the nocturnal niche depth $H$, the preferred nighttime position of the organisms. The red curve shows the smoothed interpolation used to compute the normalized mean position and excess kurtosis of the distribution. Examples of:
(a) {\it Surface Accumulation}  ($Pe = 100,\; W = 0.1$), 
(b) {\it Shallow DVM} ($Pe = 0.1,\; W = 0.1$), 
(c) {\it Deep DVM} ($Pe = 10,\; W = 0.8$, and (d) {\it  Sinking} ($Pe = 10,\; W = 1.2$)}. 
\label{fig4}
\end{figure}

\noindent\textbf{Surface Accumulation}. This regime occurs when organisms exhibit high persistent orientation and therefore asymmetric motility that favors upward swimming, as in the case illustrated in Fig.~\ref{fig3}a for $Pe = 100$ and $W = 0.1$. Under these conditions, organisms actively explore the upper water column but remain confined between the surface and the prescribed nocturnal niche ($z/H = 1$; Fig.~\ref{fig1}b). The resulting vertical distribution reflects a balance between upward dispersion and light-driven stabilization at the niche isolume. The probability density function (PDF) of the vertical concentration profile $n(z)$, over a four-day cycle,  (Fig.~\ref{fig4}a) reveals a dominant peak near $z/H = 1$, a secondary peak at the surface ($z = 0$), and a relatively flat concentration in between—evidence of continuous vertical movement within this confined layer (\textcolor{blue}{Supplementary Video~S1}).

\noindent\textbf{Shallow DVM}. 
In this regime, characterized by high vertical motility asymmetry favoring upward motion ($W\lesssim 0.25$), organisms rapidly accumulate at the nocturnal niche depth ($z/H=1$) but lack the swimming capacity to effectively track diel variations in light. As a result, the vertical concentration field $C_{h}(z,t)$ exhibits limited vertical mobility, constrained to shallow dives (Fig.~\ref{fig3}b and \textcolor{blue}{Supplementary Video~S2}). The PDF $n(z)$ (Fig.~\ref{fig4}b) displays a narrowly confined peak—a thin layer with negligible vertical dispersion—indicating strong localization and minimal environmental responsiveness.

\noindent\textbf{Deep DVM}. This regime represents a canonical pattern of vertical migration observed in natural systems. It is achieved by organisms that have well-balanced upward and downward vertical swimming capabilities, with $W\approx 1$. As an example, Fig.~\ref{fig3}c (\textcolor{blue}{Supplementary Video~S3}) shows organisms that respond effectively to changing light intensities, resulting in synchronized upward and downward migrations that follow the daily cycle. The vertical concentration field $C_{h}(z,t)$ exhibits a periodic structure, reflecting coordinated population movement in phase with solar forcing. The corresponding PDF $n(z)$ (Fig.~\ref{fig4}c) features a sharp peak at the nocturnal niche and an extended tail reaching into deeper waters, capturing the full amplitude of the migratory path. 

\begin{figure*}[ht!]
    \centering 
\includegraphics[width=1\textwidth]{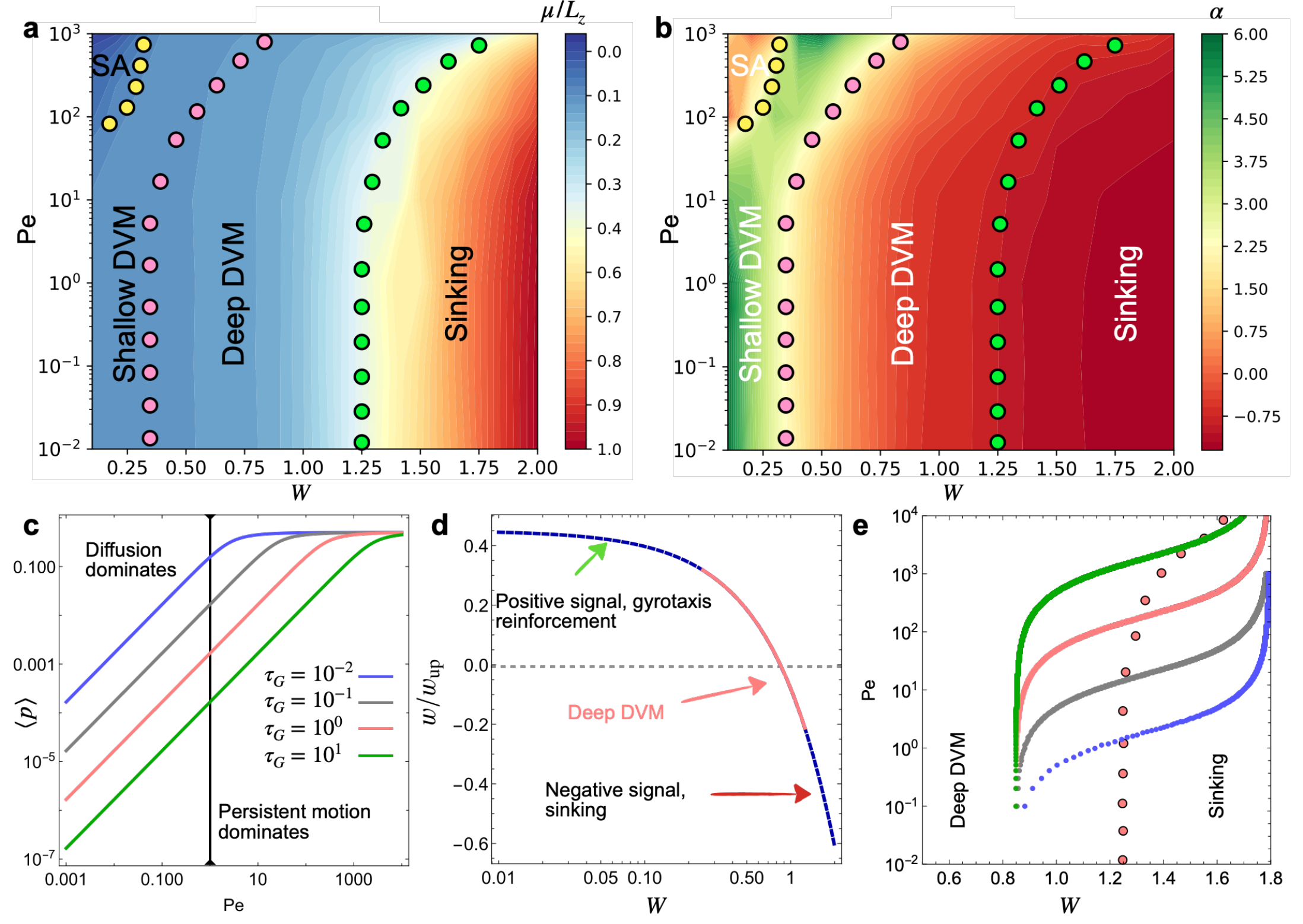}	
    \caption{\textbf{Phase diagram and swimming regimes of organisms}. (a)-(b): Phase diagrams of the normalized average $\mu/L_z$ and the excess kurtosis $\alpha$ of the time-averaged vertical distribution, over the different simulations, mapped as functions of the Péclet number $Pe$ and the vertical swimming asymmetry ratio $W$. Colored dots delineate regime boundaries: yellow indicates the transition from {\it Surface Accumulation} to trapping; pink marks the boundary between trapping and diel vertical migration; green defines the transition to the {\it Sinking} regime. (c) Mean vertical orientation $\langle p \rangle$ as a function of $Pe$ and varying gyrotactic reorientation timescale $\tau_G$: blue ($ 10^{-2}$), gray ($10^{-1}$), pink ($10^0$), and green ($10^1$). The pink curve represents the configuration used in the present model. (d) Instantaneous ratio of vertical swimming speed to upward velocity, $w/w_{\rm up}$, evaluated at depth $z = H$ and time $t$, shown as a function of $W$. Green arrow indicates the conditions under which diel vertical migration emerges; the red arrow marks the transition to {\it Sinking}. (e) Critical value of $W(Pe)$ at which the vertical velocity statistically becomes zero ($w = 0$), for different values of $\tau_G$: blue ($10^{-2}$), gray ($10^{-1}$), pink ($10^0$), and green ($10^1$).} 
\label{fig5}
\end{figure*}

\noindent\textbf{Sinking.}
This regime arises when organisms remain sensitive to light cues but are unable to track the upward motion of the niche isolume, resulting in a net downward drift. It occurs when motility is persistent but vertically biased, as in the case illustrated in Fig.~\ref{fig3}d (\textcolor{blue}{Supplementary Video~S4}) for $Pe = 10$ and $W = 1.2$. The elevated downward swimming speed overwhelms phototactic reorientation, causing organisms concentration $C_{h}(z,t)$ to sink progressively over time. Although they respond to diel light cycles, the inability to rapidly reverse direction prevents vertical recovery. The PDF $n(z)$ (Fig.~\ref{fig5}d) reflects this behavior, showing a broad, uniform distribution below the nocturnal niche.

\noindent\textbf{Mapping and bounding collective patterns}

\noindent To classify the four behavioral regimes across the $(Pe, W)$ parameter space, we evaluated the PDF $n(z)$ for all simulation cases listed in Supplementary Table~S1 (examples shown in Fig.~\ref{fig4}). From each distribution, we computed two statistical descriptors: the normalized average, $\mu/L_z$, representing the mean vertical position of the population, and the excess kurtosis, $\alpha$, which quantifies the peakedness and tail weight, thereby reflecting the extent of vertical spread. These metrics are mapped over the $(Pe, W)$ domain in Fig.~\ref{fig5}a–b (\textcolor{blue}{Methods}).

\noindent\textbf{Swarm's center of mass.} 
The normalized mean vertical position, $\mu/L_z$, captures how the population center shifts across the $(Pe, W)$ parameter space. As shown in Fig.~\ref{fig5}a, low values (blue) indicate {\it Surface Accumulation} (SA), while high values (red) reflect deeper aggregation. The isolume zone, where light intensity ranges from 50\% to 85\%, corresponds to $0.1 \leq \mu/L_z < 0.35$. Values below this range ($\mu/L_z < 0.1$) reflect accumulation in high-light surface layers, typical of the SA regime. Values above $0.5$ correspond to {\it Sinking}, with the population concentrated in deeper, darker waters (Fig.~\ref{fig4}d).

\noindent\textbf{Swarm's vertical distribution.}
The kurtosis $\alpha$ characterizes the shape of the vertical distribution. Its variation through the $(Pe, W)$ parameter space (Fig.~\ref{fig5}b) shows that high values ($\alpha \geq 3$) mark sharply peaked profiles, indicating trapping near the niche isolume. Intermediate values ($0 < \alpha < 3$) correspond to peaked distributions with extended tails, typical of SA and deep DVM. Negative or near-zero values ($\alpha \leq 0$) indicate broad, flat distributions with long tails, consistent with {\it Sinking} or deep diffusion into darker waters.\\
\noindent\textbf{Competition between stochastic vs directed motility}.

\noindent Our results show that when $Pe< 1$, where stochastic (Brownian) reorientation dominates over directed swimming, collective dynamics are largely insensitive to $Pe$ but strongly shaped by the vertical velocity ratio $W = w_{\rm down} / w_{\rm up}$. As $W$ increases, three distinct behavioral regimes emerge: shallow DVM, deep DVM, and {\it Sinking} (Fig.~\ref{fig5}a–b). In the absence of light cues, the orientation distribution of gyrotactic swimmers is governed by the Fokker–Planck equation \cite{kessler1986individual}, where azimuthal orientations are uniformly distributed and polar alignment with gravity is given by $\langle p \rangle = \coth(\lambda) - 1/\lambda$. Here, $\lambda = Pe\;[\sigma / (\tau_G u_0)]$ is a dimensionless parameter, proportional to $Pe$, comparing the gyrotactic reorientation timescale $\tau_G$ to the persistence time of the swimmer. We compute $\langle p \rangle$ as a function of $Pe$ for several values of $\tau_G$ ($10^{-2}$ to $10^1$), shown as color-coded curves from blue to green in Fig.~\ref{fig5}c. For $\tau_G = 1$ (pink curve), gyrotaxis dominates at $Pe> 10$, causing swimmers to align upward against gravity. This alignment supports the emergence of the {\it Surface Accumulation} regime within a narrow range of low $W$ (Fig.~\ref{fig3}b). More broadly, when $W < 1$, upward swimming enhances gyrotactic alignment, promoting both {\it Surface Accumulation} and {\it shallow DVM} dynamics (Fig.~\ref{fig3}a–b). In contrast, we expect this regime to vanish in the absence of gyrotaxis.

For $W > 1$, downward swimming dominates, driving organisms to dive deeper into the water column. At low-$Pe$ ($<1$), the stochastic reorientation overwhelms directional swimming, resulting in a progressive {\it Sinking} (Fig.~\ref{fig3}d). Conversely, at high-$Pe$ ($>100$), strong directional persistence allows organisms to maintain alignment with the target isolume despite the downward bias, thereby sustaining deep DVM.

Intriguingly, the deep DVM regime emerges from a finely tuned balance between opposing vertical tendencies. This raises a fundamental question: how can such a dynamic equilibrium arise from simple, local motility rules? Clues are inferred from examining how the swimming strategy modulates the organism's vertical velocity as a function of the velocity ratio $W$,
\begin{equation}
    w(z,t)/w_{\rm up}=1+\frac{ \operatorname{tanh}(A_1) +\operatorname{tanh}(A_2)}{2} - \frac{W}{2},
\label{vz}
\end{equation}
displayed in Fig.~\ref{fig5}d. This relationship reveals a positive feedback mechanism: for $W < 0.1$, the vertical velocity is enhanced, favoring upward movement and increasing the likelihood of {\it Surface Accumulation} and trapping. As $W$ increases, the net vertical velocity decreases, eventually becoming negative, promoting downward drifting. This transition suppresses directed migration and amplifies diffusion-dominated behavior, reinforcing the {\it Sinking} of organisms

From Fig.~\ref{fig5}, we identify the deep DVM regime within the interval $0.3 \leq W \leq 1.25$ at low Péclet numbers, corresponding to a transition zone where the light signal induces a switch from upward to downward vertical swimming. In this region, the magnitude of the vertical velocity satisfies $|w|/w_{up} \sim 0.4$. This delicate balance reveals the organism's capacity to regulate upward and downward swimming, which in turn may involve the ability to perform sharp reorientations. Such behavior has been observed in phytoplankton communities like \textit{H. akashiwo}~\cite{zeng2022sharp}, which exhibit brief transitions where organisms adjust their orientation to track changes in the niche isolume depth. That is, they can orient upward while downward swimming or orient downward while swimming toward the surface (\textcolor{blue}{Supplementary Text, Fig.~S4}).

To understand this mechanism, we hypothesize that the organism sets this behavior that we called {\it floating} when its vertical position reaches a steady state near the isolume, i.e., when $w = 0$, neglecting swimmer-swimmer hard-core interactions in Eq.~\ref{eq1}. This condition yields the following approximation:
\begin{equation}
    W \approx \left\{ 1 + 2\frac{u_0}{w_{\text{up}}} \left[ \coth(\lambda) - \frac{1}{2\lambda} \right] \right\}.
    \label{wpe}
\end{equation}
This functional relationship for $W$ is plotted in terms of $Pe$ in Fig.~\ref{fig5}e for different gyrotactic timescales $\tau_G = 10^{-2}, 10^{-1}, 10^{0}, 10^{1}$, represented by blue to green dashed lines, respectively. We observe that the condition $w = 0$ defines a region in which DVM can emerge. Interestingly, the smallest value of $W$ for which this condition is satisfied is $W = 1$, which appears independent of the gyrotactic timescale $\tau_G$; this limit is shifted slightly to the right in our numerical results, shown as a dotted pink line. This sets a boundary for the {\it Sinking} regime, with large $\tau_G$ (strong gyrotactic reorientation) shrinking the DVM region (see blue dashed curve) to smaller Péclet numbers.\\

\noindent{\fontsize{12}{14}{\textbf{Discussion}}}

\noindent Diel vertical migration (DVM) is the largest synchronized movement of biomass on the planet, playing a pivotal role in aquatic food webs and biogeochemical cycling \cite{bianchi2013diel,bianchi2013intensification}. While classical continuum models have captured its large-scale patterns in oceans and lakes \cite{andersen1991model,richards1996diel,bianchi2016global,piton2025identifying}, laboratory studies have revealed complex organismal mechanisms that underpin this collective phenomenon and their capacity to enhance mixing in aquatic systems \cite{noss2012zooplankton,houghton2018vertically,noto2023simple,mohebbi2024measurements}.

Here, we introduce an agent-based modeling (ABM) framework and show how macroscale DVM patterns can arise from simple, local behavioral rules in photokinetic swimmers navigating a dynamic light field. Each organism is modeled as an active gyrotactic Brownian particle that adjusts its swimming in response to local light intensity and gradients. Remarkably, we find that population-scale dynamics are governed by just two dimensionless parameters: the Péclet number ($Pe$), representing the persistence of directed motion relative to rotational diffusion, and the vertical velocity ratio ($W = w_{\text{down}} / w_{\text{up}}$), capturing asymmetry in photokinetic response across a critical isolume.

Brownian dynamics simulations (\textcolor{blue}{Methods}) across a broad $(Pe, W)$ parameter space reveal four robust collective regimes: (i) {\it Surface Accumulation}, where organisms remain near the upper water column; (ii) {\it Shallow DVM}, with tightly confined vertical oscillations; (iii) {\it Deep DVM}, characterized by pronounced excursions synchronized with isolume motion; and (iv) {\it Sinking}, where populations drift irreversibly into deeper waters. These regimes emerge from nonlinear interactions among rotational noise, gyrotactic alignment, and light-sensitive speed modulation. 

In the {\it Surface Accumulation} regime, swimmers disperse between the free-surface and the nocturnal niche, making short, asynchronous excursions during the daytime. This accumulation results from the reinforcement of the gyrotactic organism’s response and swimming strategy at small $W$, combined with the non-penetration boundary condition imposed at the free surface. Such behavior is observed in phytoplankton like dinoflagellate species \cite{zheng2023dinoflagellate} and small zooplankton rotifers like {\it Keratella cochlearis} \cite{obertegger2008multifactorial,perga2020rotiferan}, which typically remain within the upper photic zone. Their confinement is often attributed to weak swimming ability and enhanced mixing from surface-layer turbulence~\cite{de2014turbulent}, where turbulence disrupts gyrotactic shear trapping and promotes vertical dispersion~\cite{li2023environmental,wu2024onset}. In contrast, {\it Shallow DVM} occurs across all the explored Péclet numbers when the vertical velocity ratio satisfies $W \lesssim 0.25$. Organisms in this regime form sharply localized thin layers and perform synchronized, yet shallow, dives during daylight hours. This behavior mirrors gyrotactic trapping observed in laminar shear flows in the absence of external directional cues \cite{zeng2022sharp,wang2021vertical,li2023environmental,wang2022gyrotactic,hoecker2012trapping} observed in species such as the dinoflagellate {\it Akashiwo sanguinea} \cite{sullivan2010coastal} and the cyanobacterium {\it Planktothrix rubescens} \cite{walsby2002light}.
  
The {\it Deep DVM} regime represents the most widespread form of collective zooplankton migration in both oceans and lakes \cite{ringelberg2010diel,bianchi2013diel,hafker2022animal,ouillon2020active,hafker2023rhythms}. While previous studies have relied on one-dimensional continuum models incorporating predator–prey dynamics and biogeochemical fluxes \cite{richards1996diel,ringelberg2010diel,bianchi2013diel,ratnarajah2023monitoring}, our work presents the first fully three-dimensional agent-based framework that reproduces canonical DVM behavior directly from organism-scale motility rules. We find that deep DVM consistently emerges when upward and downward swimming speeds are balanced, with vertical velocity ratios clustering around $W \approx 1$. This balance yields robust diel excursions across a broad range of Péclet numbers ($Pe$), as individuals synchronize their directional switching with the daily movement of the isolume. Such behavior mirrors that observed in laboratory experiments for phytoplankton species like  {\it Heterosigma akashiwo} \cite{zeng2022sharp} and ``model'' zooplankton species like {\it Daphnia magna} \cite{dodson1997individual,noss2014direct,noto2023simple} and {\it Artemia Salina} \cite{co1998swimming,wilhelmus2014observations,houghton2018vertically}. Beyond this delicate balance lies the {\it Sinking} regime, which marks a clear boundary in the DVM phase space (Fig.~\ref{fig5}a–b). As $W$ increases beyond unity, organisms exhibit persistent downward motion and lose the ability to track the rising isolume after midday. This light-decoupled behavior occurs across all $Pe$ values and signals a collapse of diel migration—where photokinetic asymmetry overwhelms orientation and drives organisms irreversibly out of the illuminated water column.

Our study sheds light on how simple, organism-scale swimming strategies can give rise to emergent, macroscale patterns in aquatic ecosystems. By bridging individual photokinetic behaviors with population-level dynamics, this framework offers a mechanistic foundation to probe biological processes critical to oceans and lakes functioning—from carbon sequestration via the biological pump to nutrient redistribution \cite{bianchi2013diel,sengupta2017phytoplankton,steinberg2017zooplankton,gorgues2019simulated}, harmful algal blooms \cite{zeng2022sharp}, and the ecological consequences of climate change \cite{hays2003review,ringelberg2010diel,steinberg2017zooplankton,ratnarajah2023monitoring}. Beyond its ecological insights, the model also provides a predictive tool to explore how light-responsive active matter systems self-organize across environmental gradients. Future directions could integrate hydrodynamic coupling \cite{durham2009disruption,monthiller2022surfing,mousavi2024efficient,ouillon2020active,mohebbi2024measurements}, multispecies interactions \cite{leptos2023phototaxis}, and adaptive behaviors under variable light regimes \cite{hafker2023rhythms}, broadening the model’s relevance across scales and systems.\\

\noindent{\fontsize{12}{14}{\textbf{Methods}}}

\noindent
We consider $N$ self-propelled spherical particles with diameter $\sigma$, performing an overdamped Langevin dynamics in an elongated three-dimensional $L_x\times L_y\times L_z$ space with position vectors $\mathbf{r}_i(t)$ and orientations $\hat{\mathbf{p}}_i(t)$.\\

\noindent\textbf{Equations of motion}. \noindent In order to describe the stochastic dynamics of each particle, we first focus on the self-propelled motion and its corresponding active fluctuations due to the internal mechanisms enabling swimming. Then, in addition to this basic motion, further contributions arise from interparticle interactions, the response to the light signal, and a reorientation torque along the $z$-axis as a result of the particle's asymmetric distribution of mass in the presence of gravity (gyrotaxis). Thus, the motion of the $i$-th particle is described by the following dynamical system:
\begin{eqnarray}
\dot{\mathbf{r}}_i&=&u_{0}\hat{\mathbf{p}}_i-\frac{\mathbf{\nabla_{r_i}}U_i}{\gamma_T}+W\mathbf{\hat{k}},  \label{eq1}\\
    \dot{\hat{\mathbf{p}}}_i&=&\frac{1}{\gamma_R}\boldsymbol{\xi}_{i,R}\times\hat{\mathbf{p}}_i+\frac{\mathbf{\hat{k}}-(\mathbf{\hat{k}}\cdot\mathbf{\hat{p}}_i)\mathbf{\hat{p}}_i}{\tau_G}.\label{eq2}
\end{eqnarray}
\newline
Let's examine each part of it. First, the velocity $\dot{\mathbf{r}}_i$ results from: the self-propulsion term $u_{0}\hat{\mathbf{p}}_i$, with $u_0$ a fixed velocity for all particles; plus the contribution from excluded volume pair interactions $U_i$, with $\gamma_T$ being the translational Stokes friction coefficient at time $t$,
\begin{equation}
    U_i=\sum_{i\neq j}U^{ij}_{\text{WCA}}(\mathbf{r}_i(t),\mathbf{r}_j(t)),\label{eq.3}
\end{equation}
where, $\mathbf{r}_j(t)$ is the position of the $j-$th particle at time $t$, and $U_{\text{WCA}}$ is the Weeks-Chandler-Andersen (WCA) pair potential, given by:
\begin{equation}\label{eq.4}
 U_{\text{WCA}}^{ij} =  \left\{\begin{array}{cc}
     \displaystyle{4 \varepsilon\left[\left(\frac{\sigma}{r_{i j}}\right)^{12}-\left(\frac{\sigma}{r_{ij}}\right)^{6}\right]}    &   r_{i j} \leq C_{r}\\
     0    & \mbox{otherwise}. 
    \end{array} \right.
\end{equation}
\newline
Here, $C_r=2^{1/6}\sigma$ is defined as the contact radius, $r_{ij}$ corresponds to the distance between the particles $i$ and $j$, defined as $|\mathbf{r}_i - \mathbf{r}_j|$, and $\varepsilon$ is an energy interaction strength. 
Finally, the third term in Eq.~\ref{eq1}, $W = W(z, t)$, represents the contribution of the swimming strategy (see Eq.~\ref{swimst} and Fig.~\ref{fig1}c) to the vertical component of the velocity in $\hat{\mathbf{k}}$. This follows the photokinetic model proposed by Richards et al.~\cite{richards1996diel}, in which the light signal directly modulates the particle's vertical velocity. This approach contrasts with other models, where phototaxis is incorporated as a light-sensitive modulation of the swimmer's orientation~\cite{dervaux2017light,vincent1996bioconvection}.

Regarding the second equation, the zooplankton orientation $\dot{\hat{\mathbf{p}}}_i$ evolves due to two contributions: (i) active fluctuations modeled as a white Gaussian noise $\boldsymbol{\xi}_R$ with zero mean, $\langle \boldsymbol{\xi}_R(t) \rangle = 0$, and delta-correlated in time, $\langle \boldsymbol{\xi}_R(t_1)\boldsymbol{\xi}_R(t_2) \rangle = D_R \delta(t_1 - t_2)$, where $D_R$ is the rotational diffusion coefficient; and (ii) a deterministic gravitational torque term~\cite{thorn2010transport,pedley1990new}, which describes the tendency of the swimmer to reorient toward the vertical direction over a characteristic timescale $\tau_G$.\\

\noindent\textbf{Nondimensional parameters and numerical simulations}. We employed standard Brownian 
dynamics methods (explicit Euler integration) to numerically solve the governing equations \eqref{eq1} and \eqref{eq2} for a population of $N$ (\textcolor{blue}{Supplementary Table~S1}) organisms moving in a three-dimensional water column of size $L_x\times L_y\times L_z$. Periodic boundary conditions were applied in the $x$-$y$ plane,  while a Dirichlet condition $w(z=0)=0$ was imposed at the free surface and an open boundary was set at the bottom $z=L_z$. To handle periodic boundary conditions, we used the minimum-image convention to compute hard-core interactions near the borders. To reduce redundant computations, we evaluated pairwise interactions exploiting symmetry (i.e., considering only $j>i$) and assigned each particle a Verlet neighbor list. The initial distribution of particles followed a Gaussian profile centered at the nocturnal isolume depth $H$, while particle orientations were initially drawn from a uniform distribution with zero mean and unit standard deviation. Under these conditions, at each time step of size $\Delta t$, the noise generated via Box-Muller transform was coupled to gyrotaxis, and particle positions were updated according to self-propulsion, pairwise interactions, and the swimming strategy, using explicit Euler integration over a total of $T$ iterations (\textcolor{blue}{Supplementary Table~S1}).

In order to generalize the system and reduce the number of free parameters, we set $\tau_G$~\cite{pedley1990new} and the particle diameter $\sigma$, as the time and length units of our problem, respectively. Environmental factors introduced a new set of parameters, which we kept fixed throughout the numerical simulations: the bulk coefficient of light spatial attenuation $\kappa_d$, the light intensity $I_0$, and the day time scale $\tau_{\text{day}}$.

We are interested in studying the effects on collective motion arising from individual response to the light signal, here accomplished by the swimming strategy, self-propulsion, and diffusive motion. Therefore we define two dimensionless parameters: the P\'eclet number $Pe=u_0/(D_R \,\sigma)$ which compares advective to diffusive time scales, and the vertical velocity ratio $W=w_{\rm down}/w_{\rm up}$, which captures the imbalance in the vertical swimming during the photokinetic response.

We performed 120 simulations varying the pair $(Pe, W)$, through the parameters $D_R$ and $w_{\rm down}$. To account for a mixed phenotype collection of organisms~\cite{hafker2023rhythms,sengupta2017phytoplankton,krishnamurthy2020scale}, we set a value $w_{\rm down}$ and we assigned to each organism a given downward velocity generated from a Poisson distribution, $P(X=k)=\frac{\Lambda^ke^{-\Lambda}}{k!}$, with $\Lambda=w_{\rm down}$ representing the mean value of the distribution. Finally, we perform a statistical analysis of the simulations, which is detailed in the \textcolor{blue}{Supplemental Information}.\\


\noindent\textbf{Operators and observables}. We utilize different operators and we estimate different physical quantities. 

The horizontal average of the organisms concentration field $C(\bm{x},t)$ -- illustrated in Fig.~\ref{fig3} -- is determined as
\begin{equation}
    C_{h}(z,t) = \frac{1}{A_{xy}} \int^{L_{y}}_{0}\int^{L_{x}}_{0}C(\bm{x},t)\,dx\,dy,
\end{equation}
where $A_{xy} = L_{x}\times L_{y}$ is the area of the horizontal domain $x$-$y$.

The time- and horizontally-averaged quantity of a three-dimensional scalar field $f(\bm{x},t)$, like the organism's vertical velocity $w(\bm{x},t)$ or concentration $c(\bm{x},t)$ is denoted as
\begin{equation}
    \langle f\rangle = \frac{1}{\Delta T}\int^{t_{f}}
_{t_{i}} \frac{1}{A_{xy}}\int^{L_{y}}_{0}\int^{L_{x}}_{0} f(\bm{x},t)\,dx\,dy\,dt,
\end{equation}
$\Delta T=t_{f}-t_{i}$ is the time window over which the time-average is performed. For instance, the time- and horizontally-averaged organisms concentration, analyzed in Fig.~\ref{fig4}, is denoted as $n(z)=\langle C\rangle$.\\

\noindent\textbf{Probability Density Function}\\  
We construct a histogram from the time-averaged vertical position of the organisms, shown as purple bins in Fig.~\ref{fig4} (see details in \textcolor{blue}{Supplementary Text}). From this histogram, we compute the corresponding probability density function (PDF) using NumPy, SciPy, and MatPlotLib in Python. This fit is plotted as a red curve in Fig.~\ref{fig4}.\\

\noindent\textbf{Statistics: Mean and Kurtosis}\\  
To construct the colormap in Fig.~\ref{fig5}, we compute the PDF for each simulation corresponding to a given pair of parameters $(Pe, W)$, and analyze its statistical properties—specifically, the mean $\mu$ and kurtosis $\alpha$. We then plot the average values of $\mu$ and $\alpha$ over the set of simulations and interpolate these quantities to generate the phase diagram.

\bibliography{references}
\vspace{0.5cm}

\noindent{\fontsize{12}{14}{\textbf{Acknowledgments}}}

\noindent F.G.-L. and J.M were supported by Fondecyt Iniciaci\'on No.\ 11220683. This research was partially supported by the supercomputing infrastructure of the NLHPC (CCSS210001). D.N. and H.N.U. were supported by start-up grant University of Pennsylvania. This project was supported by the PURM Program 2022 University of Pennsylvania: ``Escaping from light — a circadian migration of active matter in aquatic systems''.\\

\noindent{\fontsize{12}{14}{\textbf{Author contributions statement}}}

\noindent F.G.-L. and H.N.U. conceived the problem and the modeling framework. D.N. developed the swimming strategy model. J.M. and P.M. developed the code. J.M. performed the simulations. All authors analyzed the results and reviewed the manuscript.\\

\noindent{\fontsize{12}{14}{\textbf{Competing interests}}}

\noindent The authors declare no competing interests.\\

\noindent{\fontsize{12}{14}{\textbf{Data availability}}}

\noindent The data to reproduce the figures is provided.




\end{document}